\documentclass[11pt]{article}

\usepackage[margin=1in]{geometry}
\usepackage{graphicx}
\usepackage[pdftex,colorlinks=true,linkcolor=blue,citecolor=blue,urlcolor=black]{hyperref}
\usepackage{amsmath, amsthm, amssymb}
\usepackage{subfigure}
\usepackage{comment}
\usepackage{url}
\usepackage{pdflscape}
\usepackage[ruled,lined,linesnumbered]{algorithm2e}
\usepackage{tikz}
\usetikzlibrary{circuits.logic.US}
\usepackage{multicol}

\newcommand{\R}{\mathbb{R}}

\newcommand{\Z}{\mathbb{Z}}

\newcommand{\ket}[1]{| #1 \rangle}

\DeclareMathOperator{\poly}{poly}

\newcommand{\be}{\begin{equation}}
\newcommand{\ee}{\end{equation}}
\newcommand{\bea}{\begin{eqnarray}}
\newcommand{\eea}{\end{eqnarray}}
\newcommand{\bes}{\begin{equation*}}
\newcommand{\ees}{\end{equation*}}
\newcommand{\beas}{\begin{eqnarray*}}
\newcommand{\eeas}{\end{eqnarray*}}

\newtheorem*{thm*}{Theorem}

\newtheorem*{lem*}{Lemma}

\DeclareMathOperator{\BPP}{\mathsf{BPP}}
\DeclareMathOperator{\BQP}{\mathsf{BQP}}

\DeclareMathOperator{\NP}{\mathsf{NP}}

\DeclareMathOperator{\Ptime}{\mathsf{P}}

\DeclareMathOperator{\QMA}{\mathsf{QMA}}

\begin{document}

\title{Quantum algorithms: an overview}
\author{Ashley Montanaro\thanks{School of Mathematics, University of Bristol, UK; {\tt ashley.montanaro@bristol.ac.uk}.}}
\maketitle

\begin{abstract}
Quantum computers are designed to outperform standard computers by running quantum algorithms. Areas in which quantum algorithms can be applied include cryptography, search and optimisation, simulation of quantum systems, and solving large systems of linear equations. Here we briefly survey some known quantum algorithms, with an emphasis on a broad overview of their applications rather than their technical details. We include a discussion of recent developments and near-term applications of quantum algorithms.
\end{abstract}



\section{Introduction}

A quantum computer is a machine designed to use quantum mechanics to do things which cannot be done by any machine based only on the laws of classical physics. Eventual applications of quantum computing range from breaking cryptographic systems to the design of new medicines. These applications are based on quantum algorithms -- algorithms which run on a quantum computer and achieve a speedup, or other efficiency improvement, over any possible classical algorithm. Although large-scale general-purpose quantum computers do not yet exist, the theory of quantum algorithms has been an active area of study for over 20 years. Here we aim to give a broad overview of quantum algorithmics, focusing on algorithms with clear applications and rigorous performance bounds, and including recent progress in the field.

Contrary to a rather widespread popular belief that quantum computers have few applications, the field of quantum algorithms has developed into an area of study large enough that a brief survey such as this cannot hope to be remotely comprehensive. Indeed, at the time of writing the ``Quantum Algorithm Zoo'' website cites 278 papers on quantum algorithms~\cite{qazoo}. There are now a number of excellent surveys about quantum algorithms~\cite{childs10b,mosca12,santha08,bacon10}, and we defer to these for details of the algorithms we cover here, and many more. In particular, we omit all discussion of {\em how} the quantum algorithms mentioned work. We will also not cover the important topics of how to actually build a quantum computer~\cite{ladd10} (in theory or in practice) and quantum error-correction~\cite{fowler12}, nor quantum communication complexity~\cite{buhrman10} or quantum Shannon theory~\cite{wilde13}.


\subsection{Measuring quantum speedup}

What does it mean to say that a quantum computer solves a problem more quickly than a classical computer? As is typical in computational complexity theory, we will generally consider asymptotic scaling of complexity measures such as runtime or space usage with problem size, rather than individual problems of a fixed size. In both the classical and quantum settings, we measure runtime by the number of elementary operations used by an algorithm. In the case of quantum computation, this can be measured using the quantum circuit model, where a quantum circuit is a sequence of elementary quantum operations called quantum gates, each applied to a small number of qubits (quantum bits). To compare the performance of algorithms, we use computer-science style notation $O(f(n))$, which should be interpreted as ``asymptotically upper-bounded by $f(n)$''.

We sometimes use basic ideas from computational complexity theory~\cite{papadimitriou94}, and in particular the notion of complexity classes, which are groupings of problems by difficulty. See Table~\ref{table:ccglossary} for informal descriptions of some important complexity classes. If a problem is said to be {\em complete} for a complexity class, this means that it is one of the ``hardest'' problems within that class: it is contained within that class, and every other problem within that class reduces to it.

\begin{table}
\begin{center}
  \begin{tabular}{|c|l|l|}
  \hline
  {\bf Class} & {\bf Informal definition}\\
   \hline $\Ptime$ & Can be solved by a deterministic classical computer in polynomial time\\
    $\BPP$ & Can be solved by a probabilistic classical computer in polynomial time\\
     $\BQP$ & Can be solved by a quantum computer in polynomial time\\
     $\NP$ & Solution can be checked by a deterministic classical computer in polynomial time\\
     $\QMA$ & Solution can be checked by a quantum computer in polynomial time\\
   \hline
  \end{tabular}
  \caption{Some computational complexity classes of importance in quantum computation. ``Polynomial time'' is short for ``in time polynomial in the input size''.}
  \label{table:ccglossary}
\end{center}
\end{table}


\section{The hidden subgroup problem and applications to cryptography}
\label{sec:crypto}

One of the first applications of quantum computers discovered was Shor's algorithm for integer factorisation~\cite{shor97}. In the factorisation problem, given an integer $N = p \times q$ for some prime numbers $p$ and $q$, our task is to determine $p$ and $q$. The best classical algorithm known (the general number field sieve) runs in time $\exp(O((\log N)^{1/3} (\log \log N)^{2/3}))$~\cite{buhler93}\footnote{In fact, this is a heuristic bound and this algorithm's worst-case runtime has not been rigorously determined; the best proven bound is somewhat higher.}, while Shor's quantum algorithm solves this problem substantially faster, in time $O((\log N)^3)$. This result might appear only of mathematical interest, were it not for the fact that the widely-used RSA public-key cryptosystem~\cite{rivest78} relies on the hardness of integer factorisation. Shor's efficient factorisation algorithm implies that this cryptosystem is insecure against attack by a large quantum computer.

As a more specific comparison than the above asymptotic runtimes, in 2010 Kleinjung et al.~\cite{kleinjung10} reported classical factorisation of a 768-bit number, using hundreds of modern computers over a period of two years, with a total computational effort of $\sim 10^{20}$ operations. A detailed analysis of one fault-tolerant quantum computing architecture~\cite{fowler12}, making reasonable assumptions about the underlying hardware, suggests that a {\em 2000-bit} number could be factorised by a quantum computer using $\sim 3 \times 10^{11}$ quantum gates, and approximately a billion qubits, running for just over a day at a clock rate of 10MHz. This is clearly beyond current technology, but does not seem unrealistic as a long-term goal.

Shor's approach to integer factorisation is based on reducing the task to a special case of a mathematical problem known as the hidden subgroup problem (HSP)~\cite{boneh95,brassard97}, then giving an efficient quantum algorithm for this problem.

\begin{center}
\fbox{\begin{minipage}{.9\textwidth}
{\bf Hidden subgroup problem.}
Let $G$ be a group and let $X$ be a set. Given the ability to evaluate a function $f:G \rightarrow X$, where $f$ is constant on the cosets of some unknown subgroup $H \le G$, and distinct on each coset, identify $H$.
\end{minipage}
}
\end{center}

Shor's algorithm solves the case $G = \Z$. Efficient solutions to the HSP for other groups $G$ turn out to imply efficient algorithms to break other cryptosystems; we summarise some important cases of the HSP and some of their corresponding cryptosystems in Table \ref{table:cryptosystems}. Two particularly interesting cases of the HSP for which polynomial-time quantum algorithms are not currently known are the dihedral and symmetric groups. A polynomial-time quantum algorithm for the former case would give an efficient algorithm for finding shortest vectors in lattices~\cite{regev04a}; an efficient quantum algorithm for the latter case would give an efficient test for isomorphism of graphs (equivalence under relabelling of vertices).

\begin{table}
\begin{center}
  \begin{tabular}{|l|l|l|l|}
  \hline
  {\bf Problem} & {\bf Group} & {\bf Complexity} & {\bf Cryptosystem} \\
  \hline Factorisation & $\Z$ & Polynomial~\cite{shor97} & RSA\\
  Discrete log & $\Z_{p-1} \times \Z_{p-1}$ & Polynomial~\cite{shor97} & Diffie-Hellman, DSA, \dots\\
  Elliptic curve discrete log & Elliptic curve & Polynomial~\cite{proos03} & ECDH, ECDSA, \dots\\
  Principal ideal & $\R$ & Polynomial~\cite{hallgren07} & Buchmann-Williams\\
  Shortest lattice vector & Dihedral group & Subexponential~\cite{kuperberg05,regev04} & NTRU, Ajtai-Dwork, \dots\\
  Graph isomorphism & Symmetric group & Exponential & $-$\\
   \hline
  \end{tabular}
  \caption{Some problems which can be expressed as hidden subgroup problems (HSPs). The table lists the time complexity of the best quantum algorithms known for the HSPs, and the cryptosystems that are (or would be) broken by polynomial-time algorithms.}
  \label{table:cryptosystems}
\end{center}
\end{table}


\section{Search and optimisation}
\label{sec:seo}

One of the most basic problems in computer science is unstructured search. This problem can be formalised as follows:

\begin{center}
\fbox{\begin{minipage}{.9\textwidth}
{\bf Unstructured search problem.}
Given the ability to evaluate a function $f:\{0,1\}^n \rightarrow \{0,1\}$, find $x$ such that $f(x)=1$, if such an $x$ exists; otherwise, output ``not found''.
\end{minipage}
}
\end{center}

It is easy to see that any classical algorithm which solves the unstructured search problem with certainty must evaluate $f$ $N=2^n$ times in the worst case. Even if we seek a randomised algorithm which succeeds, say, with probability $1/2$ in the worst case, the number of evaluations required is of order $N$. However, remarkably, there is a quantum algorithm due to Grover~\cite{grover97} which solves this problem using $O(\sqrt{N})$ evaluations of $f$ in the worst case\footnote{Grover's original algorithm solved the special case where the solution is unique; the extension to multiple solutions came slightly later~\cite{boyer98}.}.
The algorithm is bounded-error; that is, it fails with probability $\epsilon$, for arbitrarily small (but fixed) $\epsilon > 0$. Although $f$ may have some kind of internal structure, Grover's algorithm does not use this at all; we say that $f$ is used as an {\em oracle} or {\em black box} in the algorithm.

Grover's algorithm can immediately be applied to any problem in the complexity class $\NP$. This class encapsulates decision problems whose solutions can be checked efficiently, in the following sense: There exists an efficient classical checking algorithm $\mathcal{A}$ such that, for any instance of the problem where the answer should be ``yes'', there is a {\em certificate} which can be input to $\mathcal{A}$ such that $\mathcal{A}$ accepts the certificate. In other words, a certificate is a proof that the answer is ``yes'', which can be checked by $\mathcal{A}$. On the other hand, for any instance where the answer should be ``no'', there should be no certificate that can make $\mathcal{A}$ accept it. The class $\NP$ encompasses many important problems involving optimisation and constraint satisfaction.

Given a problem in $\NP$ that has a certificate of length $m$, by applying Grover's algorithm to $\mathcal{A}$ and searching over all possible certificates, we obtain an algorithm which uses time $O(2^{m/2} \poly(m))$, rather than the $O(2^m \poly(m))$ used by classical exhaustive search over all certificates. This (nearly) quadratic speedup is less dramatic than the super-polynomial speedup achieved by Shor's algorithm, but can still be rather substantial. Indeed, if the quantum computer runs at approximately the same clock speed as the classical computer, this implies that problem instances of approximately twice the size can be solved in a comparable amount of time.

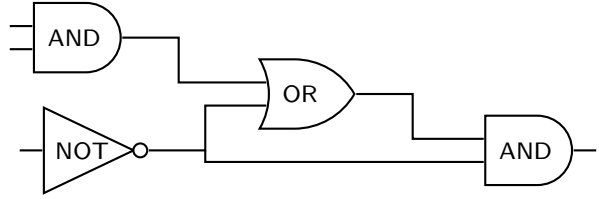
\begin{figure}
\begin{center}
\begin{tikzpicture}[thick,circuit logic US,scale=1.5]
\node[and gate,inputs={nn}] (A) at (0,0) {\tiny \sf AND};
\node[or gate,inputs={nn}] (B) at (2,-0.5) {\tiny \sf OR};
\node[not gate] (C) at (0,-1) {\tiny \hspace{0.05cm} \sf NOT\hspace{0.05cm}};
\node[and gate,inputs={nn}] (D) at (4,-1) {\tiny \sf AND};

\draw (A.input 1) -- +(-0.2,0); 
\draw (A.input 2) -- +(-0.2,0); 
\draw (C.input) -- +(-0.2,0); 
\draw (A.output) -- +(0.5,0) |- (B.input 1);
\draw (C.output) -- +(0.5,0) |- (B.input 2);
\draw (B.output) -- +(0.5,0) |- (D.input 1);
\draw (D.output) -- +(0.2,0); 
\path (C.output) -- +(0.5,0);
\draw ($(C.output)+(0.5,0)$) |- (D.input 2);
\end{tikzpicture}
\end{center}
\caption{An instance of the Circuit SAT problem. The answer should be ``yes'' as there exists an input to the circuit such that the output is 1.}
\label{fig:circuitsat}
\end{figure}

As a prototypical example of this, consider the fundamental $\NP$-complete circuit satisfiability problem (Circuit SAT), which is illustrated in Figure \ref{fig:circuitsat}. An instance of this problem is a description of an electronic circuit comprising AND, OR and NOT gates which takes $n$ bits as input and produces 1 bit of output. The task is to determine whether there exists an input to the circuit such that the output is 1. Algorithms for Circuit SAT can be used to solve a plethora of problems related to electronic circuits; examples include design automation, circuit equivalence and model checking~\cite{prasad05}. The best classical algorithms known for Circuit SAT run in worst-case time of order $2^n$ for $n$ input variables, i.e.\ not significantly faster than exhaustive search~\cite{williams13}. By applying Grover's algorithm to the function $f(x)$ which evaluates the circuit on input $x \in \{0,1\}^n$, we immediately obtain a runtime of $O(2^{n/2} \poly(n))$, where the $\poly(n)$ comes from the time taken to evaluate the circuit on a given input.

\subsection{Amplitude amplification}

Grover's algorithm speeds up the na\"ive classical algorithm for unstructured search. Quantum algorithms can also accelerate more complicated classical algorithms.
\begin{center}
\fbox{\begin{minipage}{.9\textwidth}
{\bf Heuristic search problem.}
Given the ability to execute a probabilistic ``guessing'' algorithm $\mathcal{A}$, and a ``checking'' function $f$, such that
\[ \Pr[\mathcal{A} \text{ outputs $w$ such that $f(w) = 1$}] = \epsilon, \]
output $w$ such that $f(w)=1$. 
\end{minipage}
}
\end{center}
One way to solve the heuristic search problem classically is simply to repeatedly run $\mathcal{A}$ and check the output each time using $f$, which would result in an average of $O(1/\epsilon)$ evaluations of $f$. However, a quantum algorithm due to Brassard et al.~\cite{brassard02} can find $w$ such that $f(w)=1$ with only $O(1/\sqrt{\epsilon})$ uses of $f$, and failure probability arbitrarily close to 0, thus achieving a quadratic speedup. This algorithm is known as {\em amplitude amplification}, by analogy with classical probability amplification. 

The unstructured search problem discussed above fits into this framework, by simply taking $\mathcal{A}$ to be the algorithm which outputs a uniformly random $n$-bit string. Further, if there are $k$ inputs $w \in \{0,1\}^n$ such that $f(w)=1$, then
\[ \Pr[\mathcal{A} \text{ outputs $w$ such that $f(w) = 1$}] = \frac{k}{N}, \]
so we can find a $w$ such that $f(w)=1$ with $O(\sqrt{N/k})$ evaluations of $f$. However, we could imagine $\mathcal{A}$ being a more complicated algorithm or heuristic targeted at a particular problem we would like to solve. For example, one of the most efficient classical algorithms known for the fundamental constraint satisfaction problem 3-SAT\footnote{The input is a boolean formula on $n$ variables written in conjunctive normal form with at most 3 variables per clause; our task is to determine whether the formula is satisfiable. For example, on input $(x_1 \vee \lnot x_2 \vee x_3) \wedge (x_2 \vee \lnot x_1)$ we should output ``yes'', as witnessed by $x_1=0$, $x_2=1$, $x_3=1$.} is randomised and runs in time $O((4/3)^n \poly(n))$~\cite{schoning99}. Amplitude amplification can be applied to this algorithm to obtain a quantum algorithm with runtime $O((4/3)^{n/2} \poly(n))$, illustrating that quantum computers can speed up non-trivial classical algorithms for $\NP$-complete problems.

An interesting future direction for quantum algorithms is finding accurate {\em approximate} solutions to optimisation problems. Recent work of Farhi, Goldstone and Gutmann~\cite{farhi14} gave the first quantum algorithm for a combinatorial task (simultaneously satisfying many linear equations of a certain form) which outperformed the best classical algorithm known in terms of accuracy; in this case, measured by the fraction of equations satisfied. This inspired a more efficient classical algorithm for the same problem~\cite{barak15}, leaving the question open of whether quantum algorithms for optimisation problems can substantially outperform the accuracy of their classical counterparts.


\subsection{Applications of Grover's algorithm and amplitude amplification}
\label{sec:groverapps}

Grover's algorithm and amplitude amplification are powerful subroutines which can be used as part of more complicated quantum algorithms, allowing quantum speedups to be obtained for many other problems. We list just a few of these speedups here.

\begin{enumerate}
\item Finding the {\bf minimum} of an unsorted list of $N$ integers (equivalently, finding the minimum of an arbitrary and initially unknown function $f:\{0,1\}^n \rightarrow \Z$). A quantum algorithm due to D\"urr and H\o yer~\cite{durr96} solves this problem with $O(\sqrt{N})$ evaluations of $f$, giving a quadratic speedup over classical algorithms. Their algorithm is based on applying Grover's algorithm to a function $g:\{0,1\}^n \rightarrow \{0,1\}$ defined by $g(x) = 1$ if and only if $f(x) < T$ for some threshold $T$. This threshold is initially random, and then updated as inputs $x$ are found such that $f(x)$ is below the threshold.

  \item Determining {\bf graph connectivity}. To determine whether a graph on $N$ vertices is connected requires time of order $N^2$ classically in the worst case. D\"urr et al.~\cite{durr04} give a quantum algorithm which solves this problem in time $O(N^{3/2})$, up to logarithmic factors, as well as efficient algorithms for some other graph-theoretic problems (strong connectivity, minimum spanning tree, shortest paths).
  
  \item {\bf Pattern matching}, a fundamental problem in text processing and bioinformatics. Here the task is to find a given pattern $P$ of length $M$ within a text $T$ of length $N$, where the pattern and the text are strings over some alphabet. Ramesh and Vinay have given a quantum algorithm~\cite{ramesh03} which solves this problem in time $O(\sqrt{N} + \sqrt{M})$, up to logarithmic factors, as compared with the best possible classical complexity $O(N + M)$. These are both worst-case time bounds, but one could also consider an average-case setting where the text and pattern are both picked at random. Here the quantum speedup is more pronounced: there is a quantum algorithm which combines amplitude amplification with ideas from the dihedral hidden subgroup problem and runs in time $O(\sqrt{N/M} 2^{O(\sqrt{\log M})})$ up to logarithmic factors, as compared with the best possible classical runtime $O(N/M + \sqrt{N})$~\cite{montanaro14}. This is a super-polynomial speedup when $M$ is large.
\end{enumerate}


\subsection{Adiabatic optimisation}

An alternative approach to quantum combinatorial optimisation is provided by the quantum adiabatic algorithm~\cite{farhi00}. 
The adiabatic algorithm can be applied to any constraint satisfaction problem (CSP) where we are given a sequence of constraints applied to some input bits, and are asked to output an assignment to the input bits which maximises the number of satisfied constraints. Many such problems are $\NP$-complete and of significant practical interest.
The basic idea behind the algorithm is physically motivated, and based around a correspondence between CSPs and physical systems. We start with a quantum state which is the uniform superposition over all possible solutions to the CSP. This is the ground (lowest energy) state of a Hamiltonian which can be prepared easily. This Hamiltonian is then gradually modified to give a new Hamiltonian whose ground state encodes the solution maximising the number of satisfied constraints. The quantum adiabatic theorem guarantees that, if this process is carried out slowly enough, the system will remain in its ground state throughout; in particular, the final state gives an optimal solution to the CSP. The key phrase here is ``slowly enough''; for some instances of CSPs on $n$ bits, the time required for this evolution might be exponential in $n$.

Unlike the algorithms described in the rest of this survey, the adiabatic algorithm lacks general, rigorous worst-case upper bounds on its runtime. Although numerical experiments can be carried out to evaluate its performance on small instances~\cite{farhi01}, this rapidly becomes infeasible for larger problems. One can construct problem instances on which the standard adiabatic algorithm provably takes exponential time~\cite{vandam01,farhi08}; however, changing the algorithm can evade some of these arguments~\cite{farhi11,choi11}. 

The adiabatic algorithm can be implemented on a universal quantum computer. However, it also lends itself to direct implementation on a physical system whose Hamiltonian can be varied smoothly between the desired initial and final Hamiltonians. The most prominent exponent of this approach is the company D-Wave Systems, Inc., which has built large machines designed to implement this algorithm~\cite{johnson11}, with the most recent such machine (``D-Wave 2X'') announced as having up to 1152 qubits. For certain instances of CSPs, these machines have been demonstrated to outperform classical solvers running on a standard computer~\cite{mcgeoch13,king15}, although the speedup (or otherwise) seems to have a rather subtle dependence on the problem instance, classical solver compared, and measure of comparison~\cite{ronnow14,king15}.

As well as the theoretical challenges to the adiabatic algorithm mentioned above, there are also some significant practical challenges faced by the D-Wave system. In particular, these machines do not remain in their ground state throughout, but are in a thermal state above absolute zero. Because of this, the algorithm actually performed has some similarities to classical simulated annealing, and is hence known as ``quantum annealing''. It is unclear at present whether a quantum speedup predicted for the adiabatic algorithm would persist in this setting.


\section{Quantum simulation}
\label{sec:qsim}

In the early days of classical computing, one of the main applications of computer technology was the simulation of physical systems\footnote{Such applications arguably go back at least as far as the Antikythera mechanism from the 2nd century BC.}. Similarly, the most important early application of quantum computers is likely to be the simulation of quantum systems~\cite{bulata09,brown10,georgescu14}. Applications of quantum simulation include quantum chemistry, superconductivity, metamaterials and high-energy physics. Indeed, one might expect that quantum simulation would help us understand any system where quantum mechanics plays a role.

The word ``simulation'' can be used to describe a number of problems, but in quantum computation is often used to mean the problem of calculating the dynamical properties of a system. This can be stated more specifically as: Given a Hamiltonian $H$ describing a physical system, and a description of an initial state $\ket{\psi}$ of that system, output some property of the state $\ket{\psi_t} = e^{-iHt} \ket{\psi}$ corresponding to evolving the system according to that Hamiltonian for time $t$. As all quantum systems obey the Schr\"odinger equation, this is a fundamentally important task; however, the exponential complexity of completely describing general quantum states suggests that it should be impossible to achieve efficiently classically, and indeed no efficient general classical algorithm for quantum simulation is known. This problem originally motivated Feynman to ask whether a {\em quantum} computer could efficiently simulate quantum mechanics~\cite{feynman82}.

A general-purpose quantum computer can indeed efficiently simulate quantum mechanics in this sense for many physically realistic cases, such as systems with locality restrictions on their interactions~\cite{lloyd96}. Given a description of a quantum state $\ket{\psi}$, a description of $H$, and a time $t$, the quantum simulation algorithm produces an approximation to the state $\ket{\psi_t}$. Measurements can then be performed on this state to determine quantities of interest about it. The algorithm runs in time polynomial in the size of the system being simulated (the number of qubits) and the desired evolution time, giving an exponential speedup over the best general classical algorithms known. However, there is still room for improvement and quantum simulation remains a topic of active research. Examples include work on increasing the accuracy of quantum simulation while retaining a fast runtime~\cite{berry15}; optimising the algorithm for particular applications such as quantum chemistry~\cite{hastings15}; and exploring applications to new areas such as quantum field theory~\cite{jordan12}.

The above, very general, approach is sometimes termed {\em digital} quantum simulation: we assume we have a large-scale, general-purpose quantum computer, and run the quantum simulation algorithm on it. By contrast, in {\em analogue} quantum simulation we mimic one physical system directly using another. That is, if we would like to simulate a system with some Hamiltonian $H$, we build another system which can be described by a Hamiltonian approximating $H$. We have gained something by doing this if the second system is easier to build, to run or to extract information from than the first. For certain systems analogue quantum simulation may be significantly easier to implement than digital quantum simulation, at the expense of being less flexible. It is therefore expected that analogue simulators outperforming their classical counterparts will be implemented first~\cite{bulata09}.


\section{Quantum walks}
\label{sec:qwalks}

In classical computer science the concept of the random walk or Markov chain is a powerful algorithmic tool, and is often applied to search and sampling problems. Quantum walks provide a similarly powerful and general framework for designing fast quantum algorithms. Just as a random walk algorithm is based on the simulated motion of a particle moving randomly within some underlying graph structure, a quantum walk is based on the simulated coherent quantum evolution of a particle moving on a graph.

Quantum walk algorithms generally take advantage of one of two ways in which quantum walks outperform random walks: faster hitting (the time taken to find a target vertex from a source vertex), and faster mixing (the time taken to spread out over all vertices after starting from one source vertex). For some graphs, hitting time of quantum walks can be exponentially less than their classical counterparts~\cite{childs02,kempe05}. The separation between quantum and classical mixing time can be quadratic, but no more than this~\cite{aharonov01} (approximately). Nevertheless, fast mixing has proven to be a very useful tool for obtaining general speedups over classical algorithms.

  \begin{figure}
  \begin{center}
  \begin{tikzpicture}[every node/.style={circle,fill=black,draw,inner sep=0.5mm},scale=2]
  \node (000) at (0,0,0) {};  \node[fill=yellow] (001) at (0,0,1) {A};  \node (010) at (0,1,0) {};  \node (011) at (0,1,1) {};
  \node (100) at (1,0,0) {};  \node (101) at (1,0,1) {};  \node[fill=yellow] (110) at (1,1,0) {B};  \node (111) at (1,1,1) {};
  \draw (000) to (001) to (011) to (010) to (110) to (100) to (000) to (010) to (011) to (111) to (101) to (100);
  \draw (111) to (110); \draw(001) to (101);
  \end{tikzpicture}
  \hfill
  \begin{tikzpicture}[every node/.style={circle,fill=black,draw,inner sep=0.5mm},xscale=0.8,yscale=0.3]
  \node[fill=yellow] (0) at (0,0) {A};
  \node (10) at (1,4) {}; \node (11) at (1,-4) {};
  \node (200) at (2,6) {}; \node (201) at (2,2) {}; \node (210) at (2,-2) {}; \node (211) at (2,-6) {};
  \node (3000) at (3,7) {}; \node (3001) at (3,5) {}; \node (3010) at (3,3) {}; \node (3011) at (3,1) {}; \node (3100) at (3,-1) {}; \node (3101) at (3,-3) {}; \node (3110) at (3,-5) {}; \node (3111) at (3,-7) {};
  \node (400) at (4,6) {}; \node (401) at (4,2) {}; \node (410) at (4,-2) {}; \node (411) at (4,-6) {};
  \node (50) at (5,4) {}; \node (51) at (5,-4) {};
  \node[fill=yellow] (6) at (6,0) {B};
  \draw (0) to (10); \draw (0) to (11);
  \draw (10) to (200); \draw (10) to (201); \draw (11) to (210); \draw (11) to (211);
  \draw (200) to (3000); \draw (200) to (3001); \draw (201) to (3010); \draw (201) to (3011); \draw (210) to (3100); \draw (210) to (3101); \draw (211) to (3110); \draw (211) to (3111);
  \draw (400) to (3000); \draw (400) to (3001); \draw (401) to (3010); \draw (401) to (3011); \draw (410) to (3100); \draw (410) to (3101); \draw (411) to (3110); \draw (411) to (3111);
  \draw (50) to (400); \draw (50) to (401); \draw (51) to (410); \draw (51) to (411);
  \draw (6) to (50); \draw (6) to (51);
  \end{tikzpicture}
  \hfill
  \begin{tikzpicture}[every node/.style={circle,fill=black,draw,inner sep=0.5mm},xscale=0.8,yscale=0.3]
  \node[fill=yellow] (0) at (0,0) {A};
  \node (10) at (1,4) {}; \node (11) at (1,-4) {};
  \node (200) at (2,6) {}; \node (201) at (2,2) {}; \node (210) at (2,-2) {}; \node (211) at (2,-6) {};
  \node (3000) at (3,7) {}; \node (3001) at (3,5) {}; \node (3010) at (3,3) {}; \node (3011) at (3,1) {}; \node (3100) at (3,-1) {}; \node (3101) at (3,-3) {}; \node (3110) at (3,-5) {}; \node (3111) at (3,-7) {};
  
  \node (3000c) at (4,7) {}; \node (3001c) at (4,5) {}; \node (3010c) at (4,3) {}; \node (3011c) at (4,1) {}; \node (3100c) at (4,-1) {}; \node (3101c) at (4,-3) {}; \node (3110c) at (4,-5) {}; \node (3111c) at (4,-7) {};
    
  \node (400) at (5,6) {}; \node (401) at (5,2) {}; \node (410) at (5,-2) {}; \node (411) at (5,-6) {};
  \node (50) at (6,4) {}; \node (51) at (6,-4) {};
  \node[fill=yellow] (6) at (7,0) {B};
  \draw (0) to (10); \draw (0) to (11);
  \draw (10) to (200); \draw (10) to (201); \draw (11) to (210); \draw (11) to (211);
  \draw (200) to (3000); \draw (200) to (3001); \draw (201) to (3010); \draw (201) to (3011); \draw (210) to (3100); \draw (210) to (3101); \draw (211) to (3110); \draw (211) to (3111);
  \draw (400) to (3000c); \draw (400) to (3001c); \draw (401) to (3010c); \draw (401) to (3011c); \draw (410) to (3100c); \draw (410) to (3101c); \draw (411) to (3110c); \draw (411) to (3111c);
  \draw (50) to (400); \draw (50) to (401); \draw (51) to (410); \draw (51) to (411);
  \draw (6) to (50); \draw (6) to (51);
  
  \draw (3000) to (3110c); \draw (3110c) to (3100); \draw (3100) to (3010c); \draw (3010c) to (3001); \draw (3001) to (3100c); \draw (3100c) to (3110); \draw (3110) to (3111c); \draw (3111c) to (3111); \draw (3111) to (3101c); \draw (3101c) to (3011); \draw (3011) to (3000c); \draw (3000c) to (3010); \draw (3010) to (3011c); \draw (3011c) to (3101); \draw (3101) to (3001c); \draw (3001c) to (3000);
  \end{tikzpicture}
  \end{center}
  \caption{Three graphs for whose natural generalisations to $N$ vertices a classical random walk requires exponentially more time than a quantum walk to reach the exit (B) from the entrance (A). However, on the first two graphs there exist efficient classical algorithms to find the exit which are not based on a random walk.}
  \label{fig:walks}
  \end{figure}
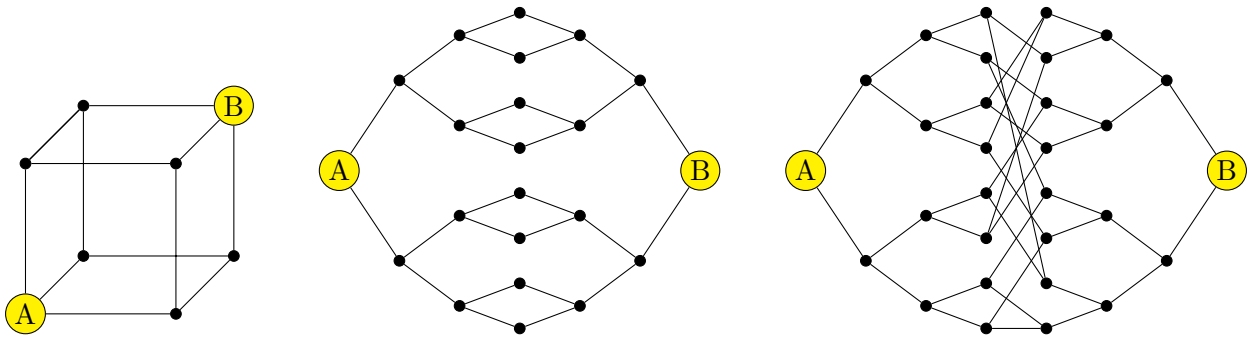
  
  Figure \ref{fig:walks} illustrates special cases of three families of graphs for which quantum walks display faster hitting than random walks: the hypercube, the ``glued trees'' graph, and the ``glued trees'' graph with a random cycle added in the middle. This third example is of particular interest because quantum walks can be shown to outperform any classical algorithm for navigating the graph, even one not based on a random walk. A continuous-time quantum walk which starts at the entrance (on the left-hand side) and runs for time $O(\log N)$ finds the exit (on the right-hand side) with probability at least $1/\poly(\log N)$. However, any classical algorithm requires time of order $N^{1/6}$ to find the exit~\cite{childs03}. Intuitively, the classical algorithm can progress quickly at first, but then gets ``stuck'' in the random part in the middle of the graph. The coherence and symmetry of the quantum walk make it essentially blind to this randomness, and it efficiently progresses from the left to the right.
 
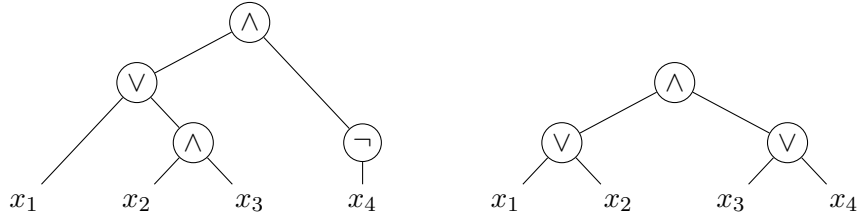
\begin{figure}[t]
  \begin{center}
  \begin{tikzpicture}[xscale=1.5,yscale=0.8]
  \node (x1) at (0,0) {$x_1$}; \node (x2) at (1,0) {$x_2$};   \node (x3) at (2,0) {$x_3$};   \node (x4) at (3,0) {$x_4$}; 
  \node[inner sep=2pt,circle,draw] (x4n) at (3,1) {$\lnot$};
  \node[inner sep=2pt,circle,draw] (x23o) at (1.5,1) {$\wedge$};
  \node[inner sep=2pt,circle,draw] (x12a) at (1,2) {$\vee$};
  \node[inner sep=2pt,circle,draw] (x12o) at (2,3) {$\wedge$};
  \draw (x4) to (x4n); \draw (x2) to (x23o); \draw (x3) to (x23o); \draw (x1) to (x12a); \draw (x23o) to (x12a);
  \draw (x12a) to (x12o); \draw (x4n) to (x12o);
  \end{tikzpicture}
  \hspace{1cm}
  \begin{tikzpicture}[xscale=1.5,yscale=0.8]
  \node (x1) at (0,0) {$x_1$}; \node (x2) at (1,0) {$x_2$};   \node (x3) at (2,0) {$x_3$};   \node (x4) at (3,0) {$x_4$}; 
  \node[inner sep=2pt,circle,draw] (x12a) at (0.5,1) {$\vee$};
  \node[inner sep=2pt,circle,draw] (x12b) at (2.5,1) {$\vee$};
  \node[inner sep=2pt,circle,draw] (x12o) at (1.5,2) {$\wedge$};
  \draw (x1) to (x12a); \draw (x2) to (x12a);  \draw (x3) to (x12b); \draw (x4) to (x12b);
  \draw (x12a) to (x12o); \draw (x12b) to (x12o);
  \end{tikzpicture}
  \end{center}
  \caption{Two boolean formulae on 4 bits. For $x_1=1$, $x_2=x_3=x_4=0$, for example, the first formula evaluates to 1 and the second to 0. The second formula is an AND-OR tree.}
  \label{fig:formula}
  \end{figure}

A possibly surprising application of quantum walks is fast evaluation of boolean formulae. A boolean formula on $N$ binary inputs $x_1,\dots,x_N$ is a tree whose internal vertices represent AND ($\wedge$), OR ($\vee$) or NOT ($\lnot$) gates applied to their child vertices, and whose $N$ leaves are labelled with the bits $x_1,\dots,x_N$. Two such formulae are illustrated in Figure \ref{fig:formula}. There is a quantum algorithm which allows any such formula to be evaluated in slightly more than $O(N^{1/2})$ operations~\cite{ambainis10b}, while it is known that for a wide class of boolean formulae, any randomised classical algorithm requires time of order $N^{0.753\dots}$ in the worst case~\cite{santha95}. The quantum algorithm is based around the use and analysis of a quantum walk on the tree graph corresponding to the formula's structure. A particularly interesting special case of the formula evaluation problem which displays a quantum speedup is evaluating AND-OR trees, which corresponds to deciding the winner of certain two-player games.

 
Quantum walks can also be used to obtain a very general speedup over classical algorithms based on Markov chains. A discrete-time Markov chain is a stochastic linear map defined in terms of its transition matrix $P$, where $P_{xy}$ is the probability of transitioning from state $x$ to state $y$. Many classical search algorithms can be expressed as simulating a Markov chain for a certain number of steps, and checking whether a transition is made to a ``marked'' element for which we are searching. A key parameter which determines the efficiency of this classical algorithm is the spectral gap $\delta$ of the Markov chain (i.e.\ the difference between the largest and second-largest eigenvalues of $P$).

There are analogous algorithms based on quantum walks which improve the dependence on $\delta$ quadratically, from $1/\delta$ to $1/\sqrt{\delta}$~\cite{ambainis03,szegedy04,magniez11}. This framework has been used to obtain quantum speedups for a variety of problems~\cite{santha08}, ranging from determining whether a list of integers are all distinct~\cite{ambainis03} to finding triangles in graphs~\cite{legall14}.


\section{Solving linear equations and related tasks}
\label{sec:lineq}

A fundamental task in mathematics, engineering and many areas of science is solving systems of linear equations. We are given an $N \times N$ matrix $A$, and a vector $b \in \R^N$, and are asked to output $x$ such that $Ax=b$. This problem can be solved in time polynomial in $N$ by straightfoward linear-algebra methods such as Gaussian elimination. Can we do better than this? This appears difficult, because even to write down the answer $x$ would require time of order $N$. The quantum algorithm of Harrow, Hassidim and Lloyd~\cite{harrow09} (HHL) for solving systems of linear equations sidesteps this issue by ``solving'' the equations in a peculiarly quantum sense: Given the ability to create the quantum state $\ket{b} = \sum_{i=1}^N b_i \ket{i}$, and access to $A$, the algorithm outputs a state approximately proportional to $\ket{x} = \sum_{i=1}^N x_i \ket{i}$. This is an $N$-dimensional quantum state which can be stored in $O(\log N)$ qubits.

The algorithm runs efficiently, assuming that the matrix $A$ satisfies some constraints. First, it should be sparse -- each row should contain at most $d$ elements, for some $d \ll N$. We should be given access to $A$ via an function to which we can pass a row number $r$ and an index $i$, with $1 \le i \le d$, and which returns the $i$'th nonzero element in the $r$'th row. Also, the condition number $\kappa =\|A^{-1}\| \|A\|$, a parameter measuring the numerical instability of $A$, should be small. Assuming these constraints, $\ket{x}$ can be approximately produced in time polynomial in $\log N$, $d$, and $\kappa$~\cite{harrow09,ambainis12b}. If $d$ and $\kappa$ are small, this is an exponential improvement on standard classical algorithms. Indeed, one can even show that achieving a similar runtime classically would imply that classical computers could efficiently simulate any polynomial-time quantum computation~\cite{harrow09}.

Of course, rather than giving as output the entirety of $x$, the algorithm produces an $N$-dimensional quantum state $\ket{x}$; to output the solution $x$ itself would then involve making many measurements to completely characterise the state, requiring time of order $N$ in general. However, we may not be interested in the entirety of the solution, but rather in some global property of it. Such properties can be determined by performing measurements on $\ket{x}$. For example, the HHL algorithm allows one to efficiently determine whether two sets of linear equations have the same solution~\cite{ambainis12b}, as well as many other simple global properties~\cite{clader13}.

The HHL algorithm is likely to find applications in settings where the matrix $A$ and the vector $b$ are generated algorithmically, rather than being written down explicitly. One such setting is the finite element method (FEM) in engineering. Recent work by Clader, Jacobs and Sprouse has shown that the HHL algorithm, when combined with a preconditioner, can be used to solve an electromagnetic scattering problem via the FEM~\cite{clader13}. The same algorithm, or closely related ideas, can also be applied to problems beyond linear equations themselves. These include solving large systems of differential equations~\cite{leyton08,berry14}, data fitting~\cite{wiebe12} and various tasks in machine learning~\cite{lloyd13}. It should be stressed that in all these cases the quantum algorithm ``solves'' these problems in the same sense as the HHL algorithm solves them: it starts with a quantum state and produces a quantum state as output. Whether this is a reasonable definition of ``solution'' depends on the application, and again may depend on whether the input is produced algorithmically or is provided explicitly as arbitrary data~\cite{aaronson15}.



\section{Few-qubit applications and experimental implementations}

Although progress in experimental quantum computation has been rapid, there is still some way to go before we have a large-scale, general-purpose quantum computer, with current implementations consisting of only a few qubits. Any quantum computation operating on at most 20-30 qubits in the standard quantum circuit model can be readily simulated on a modern classical computer. Therefore, existing implementations of quantum algorithms should usually be seen as proofs of principle rather than demonstrating genuine speedups over the classical state-of-the-art. In Table~\ref{tab:implementations} we highlight some experimental implementations of algorithms discussed here, focusing on the largest problem sizes considered thus far\footnote{Although note that one has to be careful when using ``problem size'' as a proxy for ``difficulty in solving on a quantum computer''~\cite{smolin13}.}.

An important algorithm omitted from this table is quantum simulation. This topic has been studied since the early days of quantum computation (with perhaps the first implementation dating from 1999~\cite{somaroo99}), and quantum simulations have now been implemented, in some form, on essentially every technological platform for quantum computing. One salient example is the use of a 6-qubit ion trap system~\cite{lanyon11} to implement general digital quantum simulation; we defer to survey papers~\cite{bulata09,georgescu14,blatt12,aspuruguzik12} for many further references. It is arguable that quantum simulations, in the sense of measuring the properties of a controllable quantum system, have {\em already} been performed which are beyond the reach of current classical simulation techniques~\cite{trotzky12}.

One application of digital quantum simulation which is currently the object of intensive study is quantum chemistry~\cite{hastings15,poulin15,wecker14}. Classical techniques for molecular simulation are currently limited to molecules with 50-70 spin orbitals~\cite{poulin15}. As each spin orbital corresponds to a qubit in the quantum simulation algorithm, a quantum computer with as few as 100 logical qubits could perform calculations beyond the reach of classical computation. The challenge in this context is optimising the simulation time; although polynomial in the number of orbitals, this initially seemed prohibitively long~\cite{wecker14}, but was rapidly improved via detailed analysis~\cite{poulin15}.

\begin{table}
\begin{center}
  \begin{tabular}{|l|l|p{6cm}|}
  \hline
  {\bf Algorithm} & {\bf Technology} & {\bf Problem solved} \\
   \hline Shor's algorithm & Bulk optics~\cite{martinlopez12} & Factorisation of 21\\
   Grover's algorithm & NMR~\cite{vandersypen00} & Unstructured search, $N=8$\\
   Quantum annealing & D-Wave 2X~\cite{king15} & Ising model on a ``Chimera'' graph with 1097 vertices\\
   HHL algorithm & Bulk optics~\cite{cai13,barz14}, NMR~\cite{pan14} & $2 \times 2$ system of linear equations\\
   \hline
  \end{tabular}
  \caption{Some proof-of-concept experimental implementations of quantum algorithms. Table only includes some ``largest'' problem instances solved thus far.}
  \label{tab:implementations}
\end{center}
\end{table}

The demonstration of quantum algorithms which outperform classical computation in the more immediate future is naturally of considerable interest. The Boson Sampling problem was designed specifically to address this~\cite{aaronson13}. Boson Sampling is the problem of sampling from the probability distribution obtained by feeding $n$ photons through a linear-optical network on $m$ modes, where $m \gg n$. This task is conjectured to be hard for a classical computer to solve~\cite{aaronson13}. However, Boson Sampling can be performed easily using linear optics, and indeed several small-scale experimental demonstrations with a few photons have already been carried out~\cite{ralph13}. Although Boson Sampling was not originally designed with practical applications in mind, subsequent work has explored connections to molecular vibrations and vibronic spectra~\cite{huh14,martinlopez15}.

One way in which quantum algorithms can be profitably applied for even very small-scale systems is ``quantum algorithmic thinking'': applying ideas from the design of quantum algorithms to physical problems. An example of this from the field of quantum metrology is the development of high-precision quantum measurement schemes based on quantum phase estimation algorithms~\cite{higgins07}.


\section{Zero-qubit applications}
\label{sec:noqc}

We finally mention some ways in which quantum computing is useful now, without the need for an actual large-scale quantum computer. These can be summarised as the application of ideas from the theory of quantum computation to other scientific and mathematical fields.

First, the field of Hamiltonian complexity aims to characterise the complexity of computing quantities of interest about quantum-mechanical systems. A prototypical example, and a fundamental task in quantum chemistry and condensed-matter physics, is the problem of approximately calculating the ground-state energy of a physical system described by a local Hamiltonian. It is now known that this problem -- along with many others -- is $\QMA$-complete~\cite{kitaev02,bookatz14}. Problems in the class $\QMA$ are those which can be efficiently solved by a quantum computer given access to a quantum ``certificate''\footnote{We imagine that the certificate is produced by an all-powerful (yet untrustworthy) wizard Merlin, and given to a polynomial-time human Arthur to check; hence Quantum Merlin-Arthur.}. Classically, if a problem is proven $\NP$-complete, this is considered good evidence that there is no efficient algorithm to solve it. Similarly, $\QMA$-complete problems are considered unlikely to have efficient quantum (or classical) algorithms. One can even go further than this, and attempt to characterise for which families of physical systems calculating ground-state energies is hard, and for which the problem is easy~\cite{schuch09,montanaro14}. Although this programme is not yet complete, it has already provided some formal justification for empirical observations in condensed-matter physics about relative hardness of these problems.

Second, using the model of quantum information as a mathematical tool can provide insight into other problems of a purely classical nature. For example, a strong lower bound on the classical communication complexity of the inner product function can be obtained based on quantum information-theoretic principles~\cite{cleve98}. Ideas from quantum computing have also been used to prove new limitations on classical data structures, codes and formulae~\cite{drucker11}.


\section{Outlook}
\label{sec:outlook}

We have described a rather large number of quantum algorithms, solving a rather large number of problems. However, one might still ask why more algorithms are not known -- and in particular, more exponential speedups?

One reason is that strong lower bounds have been proven on the power of quantum computation in the query complexity model, where one considers only the number of queries to the input as the measure of complexity. For example, the complexity achieved by Grover's algorithm cannot be improved by even one query while maintaining the same success probability~\cite{zalka99}. More generally, in order to achieve an exponential speedup over classical computation in the query complexity model there has to be a promise on the input, i.e.\ some possible inputs must be disallowed~\cite{beals01}. This is one reason behind the success of quantum algorithms in cryptography: the existence of hidden problem structure which quantum computers can exploit in ways that classical computers cannot. Finding such hidden structure in other problems of practical interest remains an important open problem.

In addition, a cynical reader might point out that known quantum algorithms are mostly based on a rather small number of quantum primitives (such as the quantum Fourier transform and quantum walks). An observation attributed to van Dam\footnote{See \url{http://dabacon.org/pontiff/?p=1291}.} provides some justification for this. It is known that any quantum circuit can be approximated using only Toffoli and Hadamard quantum gates~\cite{shi03}. The first of these is a purely classical gate, and the second is equivalent to the Fourier transform over the group $\Z_2$. Thus any quantum algorithm whatsoever can be expressed as the use of quantum Fourier transforms interspersed with classical processing! However, the intuition behind the quantum algorithms described above is much more varied than this observation would suggest. The inspiration for other quantum algorithms, not discussed here, includes topological quantum field theory~\cite{freedman02}; connections between quantum circuits and spin models~\cite{delascuevas11}; the Elitzur-Vaidman quantum bomb tester~\cite{lin15}; and directly solving the semidefinite programming problem characterising quantum query complexity~\cite{reichardt09,belovs14}.

As well as the development of new quantum algorithms, an important direction for future research seems to be the application of known quantum algorithms (and algorithmic primitives) to new problem areas. This is likely to require significant input from, and communication with, practitioners in other fields.


\subsection*{Acknowledgements}

This work was supported by the UK EPSRC under Early Career Fellowship EP/L021005/1. Thanks to many people including Patrick Birchall, Steve Brierley, Aram Harrow and Tom Wong for comments which improved previous versions of the paper.

\begin{multicols}{2}
\small

\bibliographystyle{plain}
\bibliography{../../thesis}

\end{multicols}

\end{document}